\documentclass[twocolumn]{article}

\title{Comments on ``Searching for a continuum limit in CDT quantum gravity"}

\author{Joshua H. Cooperman \\ \emph{Physics Program, Bard College, Annandale-on-Hudson, New York 12504}}

\usepackage{float}
\usepackage{multicol}
\usepackage{subfigure}
\usepackage{tabularx}
\usepackage{booktabs}
\usepackage{graphicx}
\usepackage{amsmath}
\usepackage{amssymb}
\usepackage{latexsym}
\usepackage{mathrsfs}
\usepackage{geometry}
\usepackage{chngcntr}
\counterwithin{figure}{section}
\counterwithin{table}{section}
\begin{document}

\twocolumn[
\begin{@twocolumnfalse}

\maketitle

\begin{abstract}
To facilitate the search for a continuum limit of causal dynamical triangulations, Ambj\o rn, Coumbe, Gizbert-Studnicki, and Jurkiewicz recently reported measurements of the lattice spacing as a function of the bare couplings. % by two methods. 
Although these authors' methods are technically sound, the conclusions that they draw from their analyses rest crucially on certain unstated assumptions. I elucidate these assumptions, and I argue that  our current understanding of causal dynamical triangulations does not entail their justification. 
\end{abstract}

\end{@twocolumnfalse}]

Causal dynamical triangulations is an approach to the quantization of classical theories of gravity based on a particular lattice regularization of the associated path integral. (See \cite{JA&AG&JJ&RL3,JHC2} for reviews.) A key question facing such an approach is %the following: does there exist 
that of the existence of a continuum limit in which physical quantities remain finite while the lattice regularization is removed. (The physical nature of this continuum limit constitutes a further---contingent---key question.) Practitioners of causal dynamical triangulations hope that a continuum limit is realized through the presence of a non-Gaussian ultraviolet fixed point, confirming Weinberg's asymptotic safety conjecture \cite{SW}. 

To determine if such a fixed point exists for a lattice-regularized theory, one first searches for second (or higher) order transitions within its phase structure. The ground state of the causal dynamical triangulations of $4$-dimensional Einstein gravity (for $3$-sphere spatial topology) possesses a rich phase structure with multiple %candidate 
phase boundaries at which higher order transitions exist %such an ultraviolet fixed point could exist 
\cite{JA&DNC&JGS&JJ2,JA&SJ&JJ&RL1,JA&SJ&JJ&RL2,DNC&JGS&JJ}. Moreover, within its so-labeled phase C, which abuts most of these higher order transitions, %candidate fixed point locations, 
the ground state exhibits physical semiclassical properties on sufficiently large scales \cite{JA&JGS&AG&JJ,JA&JGS&AG&JJ2,JA&AG&JJ&RL1,JA&AG&JJ&RL2,JA&AG&JJ&RL3,JA&JJ&RL4,JA&JJ&RL5,JA&AG&JJ&RL&JGS&TT,RK}. Having located second (or higher) order transitions, one next performs a renormalization group analysis. If an ultraviolet fixed point exists along any of these transitions, then there exist renormalization group trajectories flowing into this fixed point along which the lattice regularization necessarily vanishes. 
%In particular, one traces renormalization group trajectories through the phase structure, seeking trajectories connected to higher order transitions. If such a trajectory an ultraviolet fixed point exists along any of these transitions, then one should be able to construct a renormalization group 
Drawing on the phenomenology of phase C, 
Ambj\o rn, G\"orlich, Kreienbuehl, Jurkiewicz, and Loll made the first attempts to locate an ultraviolet fixed point and establish a continuum limit \cite{JA&AG&AK&JJ&RL}. Cooperman subsequently proposed a related but distinct renormalization group scheme %for causal dynamical triangulations 
\cite{JHC}. In the recent paper ``Searching for a continuum limit in CDT quantum gravity", Ambj\o rn, Coumbe, Gizbert-Studnicki, and Jurkiewicz (ACGSJ) seek to continue this search \cite{JA&DNC&JGS&JJ}. 

ACGSJ %state as their aim the determination of
aim to determine the dependence of the lattice spacing on the bare couplings within phase C. They foresee this information guiding the search for a continuum limit as discussed in the previous paragraph: if there exists an ultraviolet fixed point within the ground state's phase structure, then the lattice spacing approaches zero along renormalization group trajectories flowing into this fixed point. %The idea being to determine the direction of renormalization group flows towards potential ultraviolet fixed points. 
To accomplish their aim, ACGSJ employ two methods, both based on established phenomenology of phase C, to ascertain the value of the lattice spacing. First, they measure and model the amplitude of perturbations %(about the ensemble average) 
in the spatial $3$-volume \cite{JA&AG&JJ&RL2,JHC}; second, they measure and scale the diffusion time dependence of the spectral dimension \cite{JA&JJ&RL7,JA&JJ&RL6,DNC&JJ,RK}. Although %I do not fault 
the techniques that they employ in these two analyses are well-founded, I question the conclusions that they draw on the basis of their analyses. Specifically, I show that their analyses' conclusions rest crucially on respective unstated assumptions. While these assumptions could possibly be (approximately) justified, I argue that currently they are not. %I begin with some general comments, and then I consider their two methods in turn. 
I now critique their two analyses in turn. I preface each critique with an observation that lays bare the respective assumption \cite{JHC3}.

The lattice spacing %$a$ 
is dimensionful. Since one can only measure dimensionless quantities---a fact underscored by computer simulations---one cannot measure the lattice spacing itself, but one can measure the ratio of the lattice spacing to another length scale. More typically, %Put differently, 
whenever one measures a length, one actually measures the ratio of that length to an established standard unit of length. Comparing two different lengths requires that one measure both lengths with respect to the same unit of length.

In their first analysis ACGSJ seem to lose sight of this last fact. Employing a technique first demonstrated in \cite{JA&AG&JJ&RL2}---modeling perturbations in the numerically measured spatial $3$-volume about its ensemble average as linear gravitational perturbations propagating on Euclidean de Sitter space---they obtain an expression, their equation (9), for the lattice spacing $\mathfrak{a}$ in units of the Planck length $\ell_{\mathrm{P}}$. This technique thus yields a value for the dimensionless ratio $\mathfrak{a}/\ell_{\mathrm{P}}$, as the previous paragraph's observation demands. % in keeping with the above observation. 
They apply this technique to two sets of ensembles of causal triangulations: the first set consisting of four ensembles characterized by different values of $\kappa_{0}$ but the same value of $\Delta$, the second set consisting of five ensembles characterized by the same value of $\kappa_{0}$ but different values of $\Delta$. ($\kappa_{0}$ and $\Delta$ are the two bare couplings.) % characterized by different values of the bare couplings. 
They analyze their results by directly comparing values of $\mathfrak{a}/\ell_{\mathrm{P}}$ across the ensembles within each set. They claim to compare values of $\mathfrak{a}$, but, of course, they do not---and cannot---perform such a comparison. %for different values of the bare couplings. 
%The basic idea of their analysis is to compare the lattice spacing at different values of the bare couplings. 

As ACGSJ have actually measured $\mathfrak{a}/\ell_{\mathrm{P}}$, they can only meaningfully compare values of $\mathfrak{a}$ for different values of the bare couplings by assuming that $\ell_{\mathrm{P}}$ has the same value for different values of the bare couplings. They neither acknowledge this fact nor invoke this assumption, so their comparison of values of $\mathfrak{a}/\ell_{\mathrm{P}}$ remains unjustified. % do not seem to recognize this simple crucial fact, they do not explicitly make this assumption, and they therefore offer no justification for why 

What justification could ACGSJ offer for the equivalence of $\ell_{\mathrm{P}}$ for different values of the bare couplings? Recall first that $\ell_{\mathrm{P}}$ is proportional to the square root of the renormalized Newton constant $G$; accordingly, %(at the scale at which one measures it). 
equivalence of $\ell_{\mathrm{P}}$ for different values of the bare couplings follows from equivalence of $G$ for different values of the bare couplings. %Accordingly, one should attempt to justify the equivalence of $G$ for different values of the bare couplings. 
Now, ignoring momentarily the difficulty of making precise the renormalization group flow of a dimensionful coupling \cite{MA&JD}, suppose that $G$ experiences a nontrivial renormalization group flow. (If its flow is trivial, then this paragraph's opening question is also trivial.) For $G$ to have the same value for different values of the bare couplings, ACGSJ's numerical measurements must probe $G$ at the same scale for different values of the bare couplings.  
%If one naively assumes that $G$ experiences a nontrivial renormalization group flow---a scenario notably difficult to make precise owing to $G$ being dimensionful \cite{}---then one does not expect $\ell_{\mathrm{P}}$ to have the same value for different values of the bare couplings %because different values of the bare couplings likely give rise to different values for the renormalized couplings. 
%Being less naive, $\ell_{\mathrm{P}}$ is determined by the renormalized value of the Newton constant measured at some particular scale. It is moreover not clear that the scale on which the Newton constant is being measured is the same for each ensemble. Indeed, in general, this scale is not the same since this scale should vary with the bare couplings at fixed number of simplices. 
At what scale do they probe $G$? They essentially measure just three quantities to determine the value of $\mathfrak{a}/\ell_{\mathrm{P}}$: the spacetime $4$-volume, the spatial $3$-volume as a function of time, and the covariance of perturbations in the spatial $3$-volume as functions of time. %As noted above, t
The spacetime $4$-volume yields the largest scale characterizing the ground state, which ACGSJ model as the de Sitter length $\ell_{\mathrm{dS}}$. %and these other two observables are characterized by fractions of the fourth root of the spacetime volume. 
The spatial $3$-volume and the covariance of perturbations in the spatial $3$-volume vary on scales between $O(1)\,\ell_{\mathrm{dS}}$ and $O(10^{-1})\,\ell_{\mathrm{dS}}$ \cite{JHC}. If $G$ is approximately constant on these (relatively) large scales, %as perturbation theory suggests \cite{},
then $G$ is approximately equivalent for different values of the bare couplings, assuming that $\ell_{\mathrm{dS}}$ has the same value for different values of the bare couplings. 

Attempting to justify the assumption of equivalence of $\ell_{\mathrm{P}}$ for different values of the bare couplings has led to the assumption of equivalence of $\ell_{\mathrm{dS}}$ for different values of the bare couplings. %The argument has thus regressed from constancy of $\ell_{\mathrm{P}}$ across different values of the bare couplings to constancy of $\ell_{\mathrm{dS}}$ across different values of the bare couplings. 
At some point this chain of assumptions terminates with the establishment of a standard unit of length. %one must choose a length scale to establish as a standard unit. 
%The basic requirement for a 
By definition, a standard unit of length has a fixed length, the constancy of which we judge to be consistent with our current scientific knowledge. For instance, metrologists have continually updated the definition of the second to reflect our continually advancing understanding of the universe \cite{CA&BG}. ACGSJ could choose $\ell_{\mathrm{P}}$ as a standard unit of length, thereby justifying their analysis. They would then be obligated to demonstrate this choice's consistency with our current understanding of the ground state's phenomenology. Alternatively, ACGSJ could choose $\ell_{\mathrm{dS}}$ as a standard unit of length in which case %In addition to demonstrating this choice's consistency, 
they would additionally be obligated to demonstrate that all of the ensembles considered have the same value of $\ell_{\mathrm{P}}$ in units of $\ell_{\mathrm{dS}}$. Cooperman further discusses these and several other possibilities for a standard unit of length \cite{JHC}. 

The preceding discussion, particularly that of the previous paragraph, illuminates the supposition that $G$ exhibits a nontrivial renormalization group flow. Since one cannot measure dimensionful quantities, one must consider the renormalization group flow of the ratio of a dimensionful coupling to another quantity of the same dimensions. %The renormalization group flow of a dimensionful coupling is difficult to define principally owing to the above observation that one cannot measure dimensionful quantities. Accordingly, one can only define the renormalization group flow of the ratio of a dimensionful coupling to another quantity of the same dimensions. 
For instance, one could consider the renormalization group flow of $\ell_{\mathrm{P}}/\ell_{\mathrm{dS}}$, but one could attribute changes in $\ell_{\mathrm{P}}/\ell_{\mathrm{dS}}$ neither to $\ell_{\mathrm{P}}$ nor to $\ell_{\mathrm{dS}}$ alone; however, if one could establish $\ell_{\mathrm{dS}}$ as a standard unit of length, then one could attribute changes in $\ell_{\mathrm{P}}/\ell_{\mathrm{dS}}$ to $\ell_{\mathrm{P}}$ alone. %ACGSJ must carefully consider the conditions under which they can consistently treat $\ell_{\mathrm{P}}$ as constant for their first analysis' conclusions to bear scrutiny.
In light of these considerations, ACGSJ must carefully examine the consistency of their comparison of values of $\mathfrak{a}/\ell_{\mathrm{P}}$ for different values of the bare couplings. 

%Employing established literature on causal dynamical triangulations, these authors model these three measurements as follows. Linear gravitational perturbations depending only on the global time coordinate propagating on the background of Euclidean de Sitter space. In this model the spacetime volume is proportional to $\ell_{\mathrm{dS}}^{4}$, $\ell_{\mathrm{dS}}$ characterizes the scale of the spatial volume as a function of time, and the amplitude of perturbations is determine by the ratio $\ell_{\mathrm{P}}/\ell_{\mathrm{dS}}$ while the length or time scale of perturbations is proportional to $\ell_{\mathrm{dS}}$. 

%Also, one can read off this analysis directly from plot X in reference X. 

%One may conceive of a 

A renormalization group trajectory is a sequence of theories all of which describe the same physics each over a different interval of scales. For instance, along a renormalization group trajectory generated by successive coarse grainings, the interval of scales evolves from $(\ell_{\mathrm{UV}},\ell_{\mathrm{IR}})$ to $(\ell_{\mathrm{UV}}+\delta\ell,\ell_{\mathrm{IR}})$ to $(\ell_{\mathrm{UV}}+2\delta\ell,\ell_{\mathrm{IR}})$ \emph{et cetera}; the ultraviolet scale $\ell_{\mathrm{UV}}$ successively increases while the infrared scale $\ell_{\mathrm{IR}}$ remains constant. % that Each of the theories along a renormalization group trajectory describe the same physics merely over different intervals of scales. 
This perspective on renormalization group trajectories has the following immediate consequence: if one computes a scale-dependent physical observable $\mathcal{O}(\ell)$ within the theory applicable on the interval of scales $(\ell_{\mathrm{UV}},\ell_{\mathrm{IR}})$ and within the theory applicable on the interval of scales $(\ell_{\mathrm{UV}}+\delta\ell,\ell_{\mathrm{IR}})$, then the results of these two computations necessarily agree on the intersection of these two intervals of scales. This last statement makes precise the sense in which all of the theories along a renormalization group trajectory describe the same physics. 

In their second analysis ACGSJ seem to lose sight of this last fact. They first numerically measure the spectral dimension $d_{s}(\sigma)$ as a function of discrete diffusion time $\sigma$, and % for each of the ensembles within the aforementioned two sets. 
%Employing a technique first demonstrated in \cite{}, %---modeling numerical measurements of the spectral dimension with a phenomenological fit function---
%they perform numerical measurements of the spectral dimension as a function of diffusion time and model the results phenomenologically
%The authors next consider measurements of the spectral dimension as a function of diffusion time for the same ensembles of causal triangulations. 
they then phenomenologically model the functional form of $d_{s}(\sigma)$ as
\begin{equation}
d_{s}^{(\mathrm{fit})}(\sigma)=a-\frac{b}{c+\sigma}
\end{equation}
for parameters $a$, $b$, and $c$ determined by a best fit of $d_{s}^{(\mathrm{fit})}(\sigma)$ to $d_{s}(\sigma)$. They apply this model to the same two sets of ensembles of causal triangulations, obtaining values of $a$, $b$, and $c$ for each ensemble. Since the continuous diffusion time has dimensions of length squared, they next consider scaling $\sigma$, which is just a dimensionless counting parameter, as follows:
\begin{equation}\label{scaling}
\sigma\longrightarrow\left(\frac{\mathfrak{a}_{\mathrm{ref}}}{\mathfrak{a}}\right)^{2}\sigma
\end{equation}
in which $\mathfrak{a}_{\mathrm{ref}}$ is the lattice spacing of the ensemble characterized by $\kappa_{0}=2.2$ and $\Delta=0.6$, and $\mathfrak{a}$ is the lattice spacing of another ensemble. (They denote the ratio $\mathfrak{a}_{\mathrm{ref}}/\mathfrak{a}$ as $1/\mathfrak{a}_{\mathrm{rel}}$.) %Accordingly,
%\begin{equation}
%d_{s}^{(\mathrm{fit})}(\sigma)=a-\frac{b}{c+\left(\frac{\mathfrak{a}_{\mathrm{ref}}}{\mathfrak{a}_{\mathrm{rel}}}\right)^{2}\sigma}.
%\end{equation}
They obtain values of $\mathfrak{a}_{\mathrm{ref}}/\mathfrak{a}$ for each ensemble by maximizing the overlap of $d_{s}^{(\mathrm{fit})}(\sigma)$ for all of the ensembles within both sets.  %what factor of $a^{2}$ must one scale $\sigma$ so as to make these several measurements of the spectral dimension overlap. 
Finally, they contend that the values of $\mathfrak{a}_{\mathrm{ref}}/\mathfrak{a}$ so obtained indicate the relative change in the lattice spacing from the reference ensemble to each of the other ensembles.  %the authors compare lattice spacings to a fixed reference, namely the lattice spacing for a particular ensemble, so they do not lose sight of this fact, but their method of comparison runs afoul. 

For what reason do ACGSJ expect their determinations of $d_{s}^{(\mathrm{fit})}(\sigma)$ for each ensemble to overlap with $\sigma$ scaled as in equation \eqref{scaling}? I maintain that their unstated reasoning proceeds as follows. If the ensembles considered all fall along a renormalization group trajectory, then numerical measurements probe the same spectral dimension on different intervals of scales. The scaling in equation \eqref{scaling} compensates for the differences in these intervals' ultraviolet scales, resulting in the several measurements of the same spectral dimension overlapping. %(for the scales diffusion times corresponding to the intersection of the relevant intervals of scales). %(Is there a further assumption about the domain of diffusion times?)
%one has reason to expect that scaling should result in the measurements overlapping. For, in this case one is probing the same spectral dimension curve merely on different intervals of scales. 
On the contrary, if the ensembles considered do not all fall along a renormalization group trajectory, then numerical measurements probe different spectral dimensions on different intervals of scales. The scaling in equation \eqref{scaling} does not account for the differences in these spectral dimensions, resulting in the several measurements of the spectral dimension not overlapping. % these different spectral dimension curves are likely not related by a simple scaling. 
%(different shapes, unless universal for different RG trajectories)

In their determinations of $\mathfrak{a}_{\mathrm{ref}}/\mathfrak{a}$, ACGSJ have therefore assumed that all of the ensembles considered fall along a renormalization group trajectory. %This assumption is necessary for their scaling analysis to make sense. 
They neither invoke this assumption nor demonstrate that these ensembles all fall along a renormalization group trajectory, so their comparison of values of $\mathfrak{a}_{\mathrm{ref}}/\mathfrak{a}$ remains unjustified. 

Cooperman discusses how one might compare numerical measurements of the spectral dimension across ensembles characterized by different bare couplings \cite{JHC}. He argues that, %numerical measurements of the spectral dimension be of the same shape 
for ensembles all falling along a renormalization group trajectory, their spectral dimensions are of the same shape if the diffusion time is appropriately scaled. %,  at the very least he expects 
%Supposing momentarily that the scaling of $\sigma$ in equation \ref{scaling} is correct, which may not be the case for the spectral dimension on sufficiently small scales,Moreover, 
Eyeballing ACGSJ's figure 4, which depicts $d_{s}^{(\mathrm{fit})}(\sigma)$ scaled as in equation \eqref{scaling} for all of the ensembles considered, the several spectral dimensions do not all appear to be of the same shape. The scaling of $\sigma$ in equation \eqref{scaling}, based on its canonical scaling dimension, may not be correct for sufficiently small diffusion times on which the spectral dimension exhibits decidedly nonclassical behavior. Alternatively, all of the ensembles considered may not fall along a renormalization group trajectory, a possibility suggested by the conjectured renormalization group trajectory depicted in ACGSJ's figure 7. % also does not suggest that a single trajectory passes through all of the ensembles considered.
With these considerations in mind, ACGSJ must carefully examine the consistency of their comparison of %the conditions under which one can consistently compare 
numerical measurements of the spectral dimension for different values of the bare couplings. % if they  For their second analysis' conclusions to bear scrutiny, 

If ACGSJ can justify the two assumptions elucidated above, %---either along the lines that I have suggested or by other means---
then their analyses will make important contributions to the search for a continuum limit of causal dynamical triangulations. In the absence of justification, the import of their conclusions is seriously in doubt. I fear that to justify these assumptions they must tackle one of the most difficult problems facing the causal dynamical triangulations approach---and indeed all approaches---to the quantization of gravity: the construction of meaningful physical observables beyond the few already identified. %known within causal dynamical triangulations. 

\emph{Acknowledgements} I thank Hal Haggard for comments on a draft of this paper.

\end{document}